\documentclass[11pt,a4paper]{article}
\usepackage[truedimen,margin=34mm]{geometry}

\usepackage{mathrsfs}
\usepackage{amssymb}
\usepackage{amsmath}
\usepackage{ascmac}
\usepackage{amsthm}
\usepackage[pdftex]{graphicx}
\usepackage[pdftex]{color}
\usepackage{booktabs}
\usepackage{setspace}
\usepackage{natbib}

\usepackage{titlesec}
\titleformat*{\section}{\centering\large\bf}
\titleformat*{\subsection}{\centering\it}

\def\ep{{\varepsilon}}

\def\beh{{\widehat \beta}}

\def\alh{{\widehat \alpha}}

\def\Var{{\rm Var}}
\def\E{{\rm E}}

\def\bX{{\text{\boldmath $X$}}}
\def\bZ{{\text{\boldmath $Z$}}}

\def\bom{{\text{\boldmath $\omega$}}}

\def\bomh{{\widehat \bom}}

\title{{\bf 
Efficient Screening of Predictive Biomarkers for Individual Treatment Selection
}}
\date{}

\begin{document}
\doublespacing
\maketitle

\vspace{-1.5cm}
\begin{center}
{\large Shonosuke Sugasawa$^{1,2*}$, Hisashi Noma$^{2,3}$}
\end{center}
\noindent
$^1$Center for Spatial Information Science, The University of Tokyo, Chiba, Japan\\
$^2$Research Center for Medical and Health Data Science, The Institute of
Statistical Mathematics, Tokyo, Japan\\
$^3$Department of Data Science, The Institute of Statistical Mathematics, Tokyo, Japan

\vspace{3cm}
\noindent
*Correspondence: Shonosuke Sugasawa\\
\hspace{1cm}Center for Spatial Information Science\\
\hspace{1cm}The University of Tokyo\\
\hspace{1cm}5-1-5 Kashiwanoha, Kashiwa, Chiba 277-8568, Japan\\
\hspace{1cm}E-mail: sugasawa@csis.u-tokyo.ac.jp

%   Abstract
\newpage
\vspace{0.5cm}
\begin{center}
{\large\bf Summary}
\end{center}
The development of molecular diagnostic tools to achieve individualized medicine requires identifying predictive biomarkers associated with subgroups of individuals who might receive beneficial or harmful effects from different available treatments. However, due to the large number of candidate biomarkers in the large-scale genetic and molecular studies, and complex relationships among clinical outcome, biomarkers and treatments, the ordinary statistical tests for the interactions between treatments and covariates have difficulties from their limited statistical powers. In this paper, we propose an efficient method for detecting predictive biomarkers. We employ weighted loss functions of \cite{Chen2017} to directly estimate individual treatment scores and propose synthetic posterior inference for effect sizes of biomarkers. We develop an empirical Bayes approach, namely, we estimate unknown hyperparameters in the prior distribution based on data. We then provide efficient screening methods for the candidate biomarkers via optimal discovery procedure with adequate control of false discovery rate. The proposed method is demonstrated in simulation studies and an application to a breast cancer clinical study in which the proposed method was shown to detect the much larger numbers of significant biomarkers than existing standard methods.


\newpage
%-------------------------------------------------------------%
%    Introduction
%-------------------------------------------------------------%
\section{Introduction}
Due to the advances of disease biology, it has been revealed that there is substantial molecular heterogeneity among individual patients in many diseases. This implies that the benefits and harms of many treatments might also be heterogeneous, and accurate molecular diagnostic methods could maximize the treatment benefits for individual patients \citep{GN2012}. To develop the molecular diagnostic tools, a key task is the identification of predictive biomarkers associated with subgroups of individuals who might receive beneficial or harmful effects from different available treatments \citep{Matsui2015,Lip2017}. It is typically explained by interactions between the treatment and candidate biomarkers, but the conventional interaction tests have substantial difficulties for these analyses because of their serious limitations of statistical powers \citep{Matsui2018}.

On the other hand, efficient estimating methods of individual treatment effects (ITE) have been also developed for large-scale genetic and molecular studies. 
Specifically, there is a growing number of literatures regarding the efficient estimation of ITE \citep[e.g.][]{Chen2017, KU2006, Lu2018, Sugasawa2019, Tian2014, Wager2018, Zhang2017} among many others.
Although these methods can effectively estimate ITE, the estimated model is typically too complicated, and it would be difficult to understand which biomarkers are actually associate with ITE. Besides, the regularization methods often provide us not only the estimates of model parameters but also selection results of biomarkers included in the estimated model \citep{Lu2013, ZZ2018}, but the regularization method do not necessarily guarantee the control of the degree of false discoveries, i.e. false discovery rate \citep[FDR;][]{BH1995}. 
Since the primary purpose of large-scale genetic and molecular studies is screening relevant candidate markers, the prioritization and assessment of the accuracy of selections are relevant tasks.

In this article, we propose an efficient method for screening predictive biomarkers associated with ITE under control of FDR. We employ the weighted loss function proposed by \cite{Chen2017} designed to directly estimate ITE score without explicitly specifying the main effect, and construct synthetic likelihood for effect sizes of candidate biomarkers. 
The synthetic likelihood is then combined with a latent semiparametric distribution of true effect sizes, which enables us derive the synthetic posterior distribution for effect sizes. 
The underlying semiparametric distribution of effect sizes can be estimated by the EM algorithm \citep{Demp1977}, and we propose the optimal discovery procedure (ODP) to detect predictive biomarkers with adequate control of FDR. 
Although most of multiple testing procedures with FDR control are based on summary statistics such as $z$-values or $p$-values for effect sizes of biomarkers \cite[e.g.][]{ST2003, Sun2007}, the proposed method more directly use data via synthetic likelihood.
Also the proposed method would be able to increase statistical power to detect relevant biomarkers by borrowing information from related biomarkers via an empirical Bayes approach. 
In fact, through simulation studies, we numerically show that the proposed method can detect much more true predictive biomarkers than standard methods. Moreover, we apply the proposed method to an observational study of breast cancer, and we found that the proposed methods were able to detect a large number of predictive biomarkers whereas the standard methods detected less than 2 even when FDR is allowed to be $20\%$.

This paper is organized as follows. In Section \ref{sec:method}, we describe the proposed method and demonstrate the optimal discovery procedure. 
In Section \ref{sec:sim}, we conducted simulation studies to confirm the effectiveness of the proposed method compared with conventional testing approaches. 
We then apply the proposed method to a dataset of a breast cancer study in Section \ref{sec:app}. 
We give some discussions in Section \ref{sec:dis}.
Finally, R-code implementing the proposed method is available at GitHub (https://github.com/sshonosuke/SB-ITS).

%-------------------------------------------------------------%
%    Method
%-------------------------------------------------------------%
\section{Proposed Method}\label{sec:method}

%   settings
\subsection{Notations and assumptions}
We adopt the notation based on the potential outcome framework in causal inference \citep{Rubin2005}.
Let $T\in \{-1,1\}$ be the treatment indicator with $T=1$ indicating a patient being treated and $T=-1$ indicating the opposite. 
Let $Y^{(T)}$ denote the potential outcome indicating the response of a patient (e.g., survival time, binary disease status) if the patient receives treatment $T$. 
In practice, only one of the potential outcomes $Y^{(1)}$ and $Y^{(-1)}$ can be observed for each patient, that is, $Y=I(T=1)Y^{(1)}+I(T=-1)Y^{(-1)}$, where $I(\cdot)$ is the indicator function.
We employ strongly ignorable assumption \citep{RR1983, Rubin2005}, that is, $T$ is independent of $(Y^{(1)},Y^{(-1)})$ given the $p$-dimensional covariates $\bX=(X_1,\ldots,X_p)^t$ of potential predictive biomarkers of the treatment (e.g., genotypes, gene expressions).
Moreover, let $\bZ$ be a vector of potentially confounding covariates with $\bX$. 
For the treatment assignment, we assume that probability of treatment assignment is a function of $\bX$ and $\bZ$, that is, $P(T=1|\bX, \bZ)=\pi(\bX, \bZ)$, where $\pi(\bX, \bZ)$ is known as propensity score, and $\pi(\bX, \bZ)=1/2$ under randomized clinical trial or $\pi(\bX, \bZ)$ needs to be estimated (e.g. via regression modeling) in observational studies. 
We first assume that $\pi(\bX, \bZ)$ is known, and provide discussions on estimating $\pi(\bX, \bZ)$ in Section \ref{sec:app}.
The observed data $\{(Y_i,T_i,\bX_i, \bZ_i), i=1,\ldots,n\}$ consists of $n$ independent identically distributed copies of $(Y,T,\bX,\bZ)$.

%   model
\subsection{Weighted loss function and synthetic posterior}\label{sec:WL}
Our goal is to detect biomarkers among $\bX$ that are associated with individual treatment effect (ITE) denoted by $\Delta(X)$, e.g. $\Delta(\bX, \bZ)=\E[Y|\bX,\bZ,T=1]-\E[Y|\bX,\bZ,T=-1]$, with adjustment of confounding with $\bZ$.
Traditional approaches for estimating $\Delta(\bX, \bZ)$ use parametric models including main effect (function of $\bX$ and $\bZ$) and interaction effect (function of $\bX$, $\bZ$ and $T$), thereby the estimation of main effect (nuisance part) may lead to inefficient estimation of the interaction effect.
To avoid the problem, we here employ the weighted loss function \citep{Chen2017} given by
\begin{equation}\label{WL}
L_n(f)=\sum_{i=1}^n\frac{M\{Y_i,T_if(\bX_i, \bZ_i)\}}{T_i\pi(\bX_i, \bZ_i)+(1-T_i)/2},
\end{equation}
where $\pi(\bX_i, \bZ_i)=P(T_i=1|\bX_i, \bZ_i)$ is the propensity score, and $M(\cdot,\cdot)$ is a loss function.
As noted in \cite{Chen2017}, the choice of $M(u,v)$ determines the measure of treatment effect. 
For example, under $M(u,v)=(u-v)^2$, the minimizer of the true risk $\E[L_n(f)]$ is $\{\E[Y_i|\bX_i, \bZ_i,T_i=1]-\E[Y_i|\bX_i, \bZ_i,T_i=-1]\}/2$, which is difference of expectations between treatment and opposite groups.

Now we consider a linear predictor for the $k$th biomarker: $f(X_k, \bZ; \alpha_k,\beta_k, \bom_k)=\alpha_k+\beta_kX_k+\bZ^t\bom_k$, where $\alpha_k$ is the biomarker-specific intercept, $\beta_k$ is the effect size of the $k$th biomarker and $\bom_k$ is a vector of regression coefficients of $\bZ$.
Then, the weighted loss function (\ref{WL}) is reduced to 
\begin{equation}\label{Syn}
\sum_{i=1}^n\frac{M(Y_i,T_i\alpha_k+T_iX_{ik}\beta_k+T_i\bZ_i^t\bom_k)}{T_i\pi(\bX_i, \bZ_i)+(1-T_i)/2}.
\end{equation}
Since our primary interest is the effect size $\beta_k$, the other parameters $\alpha_k$ and $\bom_k$ are nuisance parameters.  
We define synthetic likelihood for the effect sizes $\beta_k$'s and the nuisance parameters based on the loss function (\ref{Syn}), which is given by
\begin{equation}\label{Syn2}
L_k(\alpha_k,\beta_k,\bom_k)=\exp\left[-\frac1A\sum_{i=1}^n\frac{M(Y_i,T_i\alpha_k+T_iX_{ik}\beta_k+T_i\bZ_i^t\bom_k)}{T_i\pi(\bX_i, \bZ_i)+(1-T_i)/2}\right],
\end{equation}
where $A$ is a scaling constant such that $A=n^{-1}\sum_{i=1}^n\{T_i\pi(\bX_i,\bZ_i)+(1-T_i)/2\}^{-1}$, and the maximizer of $L_k$ is the same as the minimizer of (\ref{Syn}).
Note that $w_i=A^{-1}\{T_i\pi(\bX_i,\bZ_i)+(1-T_i)/2\}^{-1}$ can be interpreted as the weight for the $i$th observation such that $\sum_{i=1}^nw_i=n$. 
Regarding the concrete form of the loss function $M(u,v)$, we simply adopt negative log-likelihood functions to make our empirical Bayes approaches valid. 
Therefore, the logarithm of the synthetic likelihood (\ref{Syn2}) can be regarded as a weighted log-likelihood.

From (\ref{Syn2}), $\beta_k$ can be interpreted as interaction effect of $T_i$ and $X_{ik}$, so that $\beta_k=0$ means that the $k$th biomarker is irrelevant to ITE; otherwise, the $k$th biomarker is a predictive biomarker.
Hence, we consider statistical testing whether $\beta_k$ is zero or not for each $k$.
To this end, we first note that $\alpha_k$ and $\bom_k$ are nuisance parameters in (\ref{Syn}).
For constructing (profile) synthetic likelihood of $\beta_k$ from (\ref{Syn}), we consider the following two methods:

\begin{itemize}
\item[1.](Plug-in method). \ \ 
Compute the point estimator $\alh_k$ and $\bomh_k$ of $\alpha_k$ and $\bom_k$, respectively, obtained by maximizing the synthetic likelihood function (\ref{Syn}), and replace $\alpha_k$ and $\bom_k$ in (\ref{Syn}) with $\alh_k$ and $\bomh_k$.  

\item[2.](Normal-approximation method). \ \ 
Compute the mode $\beh_k$ and the inverse value of Hessian at the mode, $s_k$, of the synthetic likelihood function (\ref{Syn}) with respect to $\beta_k$, and define the synthetic likelihood function of $\beta_k$ as $\phi(\beh_k; \beta_k,s_k)$.
\end{itemize}

The synthetic likelihood of $\beta_k$ will be denoted by $\text{PL}_k(\beta_k)$.
Since information from $n$ individuals is summarized in $\beh_k$ and $s_k$, the normal approximation method have computational advantage compared to the plug-in method which needs to compute summation over $n$ individuals.
However, the normal approximation might be poor when $n$ is not large, which may lead to less power than the plug-in method.

We consider the multiple hypothesis tests, H$_0: \beta_k=0$ vs H$_1:\beta_k\neq 0$.
In order to express the null and non-null biomarkers, we introduce the following latent structure for the effect sizes:  
\begin{equation}\label{prior}
G(\beta_k)=\pi g_0(\beta_k)+(1-\pi)g_1(\beta_k),
\end{equation}
where $\pi$ is the prior probability of being null, that is, $\pi=P(\beta_k=0)$, and the functions $g_0$ and $g_1$ represent the distributions of the null and non-null biomarkers, respectively.
For null biomarkers, we use the one-point distribution on $0$, that is, $g_0(\cdot)=\delta_0(\cdot)$, where $\delta_c(x)$ represent the one-point distribution on $x=c$.
The form of $g_1$ is not specified, so that the latent structure of $\beta_k$ can be seen as a semiparametric model where a nonparametric distribution is assumed for non-null biomarkers. 
Combined with the profile likelihood ${\rm PL}_k$ and the prior (\ref{prior}), we define the following synthetic posterior distribution of $\beta_k$:
\begin{equation}\label{pos}
\begin{split}
G(\beta_k|{\rm Data})
&=\frac{\text{PL}_k(\beta_k)G(\beta_k;\pi,g_1)}{\int \text{PL}_k(\beta_k)G(\beta_k;\pi,g_1){\rm d}\beta_k}\\
&=\frac{\pi \text{PL}_k(\beta_k)g_0(\beta_k)+(1-\pi) \text{PL}_k(\beta_k)g_1(\beta_k)}
{\pi \text{PL}_k(0)+(1-\pi) \int\text{PL}_k(\beta_k)g_1(\beta_k){\rm d}\beta_k}, 
\end{split}
\end{equation}
which are independent for $k=1,\ldots,p$.

The synthetic posterior distribution (\ref{pos}) have hyperparameters $\pi$ and $g_1$.
We consider an empirical Bayes approach, that is, we estimate these parameters from the synthetic marginal likelihood:
\begin{equation}\label{like}
\begin{split}
\prod_{k=1}^p&\int \text{PL}_k(\beta_k)G(\beta_k;\pi,g_1){\rm d}\beta_k\\
&=\prod_{k=1}^p\left\{\pi \text{PL}_k(0)+(1-\pi) \int\text{PL}_k(\beta_k)g_1(\beta_k){\rm d}\beta_k\right\}.
\end{split}
\end{equation}
We employ the smoothing-by-roughening approach \citep{SL1999} in which the nonparametric estimate of $g_1$ is supported by fixed discrete mass points, that is, we approximate $g_1(x)$ as $\sum_{\ell=1}^Lp_{\ell}\delta_{a_\ell}(x)$ with mixing probabilities $p_{\ell}$'s such that $\sum_{\ell=1}^Lp_\ell=1$, and fixed knots $a_1,\ldots,a_L$.
Then, the above synthetic likelihood can be efficiently maximized by an EM algorithm \citep{Demp1977} whose details are provided in Appendix.
Note that the mass probabilities on $a_1,\ldots,a_L$ and $0$ are $(1-\pi)p_1,\ldots,(1-\pi)p_L$ and $\pi$, respectively, which could be directly estimated via a simpler EM algorithm.
However, biomarkers-specific indices directly extracted from the proposed EM algorithm are useful for the optimal discovery procedure described in the subsequent section.

Regarding the selection of knots, some guidance is given in \cite{SL1999}, which recommends using sufficiently large range of knots to cover the main part of the true underlying distribution.
We here suggest a simple data-dependent method.
Let $\underline{\beta}$ and $\overline{\beta}$ be the minimum and maximum values of the point estimates of effect sizes $\{\beh_1,\ldots,\beh_p\}$.
Then, we set knots $a_1,\ldots,a_L$ as equally spaced points from $\underline{\beta}$ to $\overline{\beta}$, that is, $a_\ell=\underline{\beta}+(\ell-1)(\overline{\beta}-\underline{\beta})/(L-1)$ with $\ell=1,\ldots,L$, where $L$ is the number of knots.
The choice of $L$ would be less sensitive to the results as long as $L$ is large to a certain extent.
We set $L=100$ as a default choice and we cary out sensitivity analysis of $L$ in our simulation studies in Section \ref{sec:sim}.

%  Index
\subsection{Biomarker-specific indices}
Some biomarker-specific indices are useful for screening biomarkers.
Let $\gamma_k$ be the indicator variable for null/non-null status for the $k$th biomarkers, such that $\gamma_k=1$ if the $k$th biomarker is non-null and $\gamma_k=0$ otherwise. 
The value of $\gamma_k$ is unknown and has the prior probability $P(\gamma_k=1)=1-\pi$.
From the synthetic posterior (\ref{pos}), we can compute the posterior probability of being non-null as 
$$
P(\gamma_k=1 | \text{Data})
=\frac{(1-\pi) \int\text{PL}_k(\beta_k)g_1(\beta_k){\rm d}\beta_k}
{\pi \text{PL}_k(0)+(1-\pi) \int\text{PL}_k(\beta_k)g_1(\beta_k){\rm d}\beta_k},
$$
where the integral can be expressed as summation over the discrete mass points used to estimate $g_1$. 
By plugging in the hyperparameter estimates of $\pi$ and $g_1$, the estimated posterior probability of being non-null can be obtained.
Other biomarker-specific indices (e.g. posterior mean of $\beta_k$) can also be estimated based on the synthetic posterior distribution (\ref{pos}).

%  ODP
\subsection{Screening biomarkers based on the optimal discovering procedure}
For ranking biomarkers, we apply the optimal discovery procedure \citep{Storey2007,SDL2007}, and use the following statistic:
$$
\text{ODS}_k=\frac{\int\text{PL}_k(\beta_k)\widehat{g}_1(\beta_k){\rm d}\beta_k}{\text{PL}_k(0)},
$$
which is the model-based version of the optimal discovery statistic \citep{Cao2009,NM2012,NM2013}.
Since the optimal discovery procedure is known to provide maximum expected number of significant biomarkers under given a certain false discovery rate \citep{NM2012}, we may efficiently detect predictive biomarkers by using the statistic $\text{ODS}_k$. 
We select a set of $\Theta(\lambda)$ of biomarkers whose ODS values are equal to or greater than $\lambda$. 
To estimate the number of false positives in the selected set, we oblate a false discovery rate (FDR) based on the fitted model \cite[e.g.][]{Mac2006}.
$$
\text{FDR}(\lambda)=\frac1{|\Theta(\lambda)|}\sum_{k\in \Theta(\lambda)}\widehat{P}(\gamma_k=0|\text{Data}),
$$
where $|\Theta(\lambda)|$ represents the size of $\Theta(\lambda)$, the number of selected biomarkers. 
Note that the summation of $\widehat{P}(\gamma_k=0|\text{Data})$ (posterior probability of being null) represents the expected number of null (false-positive) biomarkers in $\Theta(\lambda)$. 
For specified value of $\lambda$, we may compute FDR, so that we may compute the reasonable value of $\lambda$ under user-specified FDR (e.g. 5\%).

%-------------------------------------------------------------%
%    Simulation
%-------------------------------------------------------------%
\section{Simulation study}\label{sec:sim}

We here assess the performance the proposed methods together with some alternative methods.
We considered two types of responses: binary and censored survival responses.
We set $n=1000$ (the number of samples) and $p=3000$ (the number of candidate biomarkers) throughout this study.
We generated the covariates $(X_1,\ldots,X_p)$ from a mean zero multivariate normal distribution with covariance matrix whose $(i,j)$-element is $(0.1)^{|i-j|}$. 
We also generated two confounding covariates $(Z_1,Z_2)$ from a bivariate normal distribution with $E[Z_1]=E[Z_2]=0.1X_1-0.1X_{10}$, $\Var(Z_1)=\Var(Z_2)=1$ and ${\rm Cov}(Z_1,Z_2)=0.2$.
The treatment indicator $T$ was indecently generated from a Bernoulli distribution with probability $\text{logit}^{-1}(0.2X_1+0.1X_2+0.1Z_1)$, where $\text{logit}^{-1}(x)=\exp(x)/(1+\exp(x))$.

We first consider the case with binary response.
We generated $n$ independent samples from the following model:
$$
Y=I\left(\sum_{k=1}^p\gamma_kX_k+\sum_{k=1}^p\delta_kX_k^2+T\sum_{k=1}^p\beta_kX_k+\xi_1TZ_1+\xi_2TZ_2+\ep>0\right),
$$ 
where $\ep\sim N(0,5^2)$. 
For the true parameter values of the main effect, we set $\gamma_1=\gamma_3=\gamma_5=0.2$, $\gamma_2=\gamma_4=\gamma_6=-0.2$, $\delta_1=\delta_3=\delta_5=0.2$, $\delta_2=\delta_4=\delta_6=-0.2$, $\xi_1=\xi_2=0.1$, and the other values of $\gamma_k$'s and $\delta_k$'s were set to $0$.
%$$
%\gamma_1=\gamma_3=\gamma_5=0.2, \ \gamma_2=\gamma_4=\gamma_6=-0.2, \ \delta_1=\delta_3=\delta_5=0.2
%$$
%$$
%\delta_2=\delta_4=\delta_6=-0.2, \ \ \  \xi_1=\xi_2=0.1
%$$
For effect sizes $\beta_k$ in the interaction term, we consider the following generating distribution
$$
f(x)=\pi\delta_0(x)+(1-\pi)0.3\phi(x; 0.2,(0.1)^2)+(1-\pi)0.7\phi(x;-0.5,(0.1)^2),
$$
so that the distribution of $\beta_k$ consists of three components, one-point distribution on $0$ representing null effect, two normal distributions with positive and negative means representing biomarkers that have positive and negative values on the individual treatment effects, respectively.
We considered two cases of null probability $\pi$ of $0.5$ and $0.8$.
For the simulated dataset, we applied the proposed ODP methods with the use of negative log-binomial likelihood function for the loss function in which $Z_1$ and $Z_2$ are adopted as adjustment covariates.
As explained in Section \ref{sec:method}, we considered two approaches for constructing the synthetic profile likelihood: the plug-in method and the normal-approximation method, which are denoted by ODP-P and ODP-N, respectively.
For estimating unknown non-null distribution, we employed the data-dependent method described in the end of Section \ref{sec:WL} and set $L=100$ as the number of knots.
We also applied ODP-N with $L=50, 150, 200$ to check sensitivity of the choice of $L$.
As conventional screening methods, we used the following standardized test statistics for the $k$th biomarkers:
\begin{equation}\label{standard}
T_k=\frac{\beh_k}{s_k}, \ \ \ \ \ 
S_k=\frac{\beh_{k1}-\beh_{k0}}{\sqrt{s_{k1}^2+s_{k0}^2}},
\end{equation}
where $\beh_{kj}$ and $s_{kj}$ are estimate and standard error based on the logistic regression with $k$th biomarker and two confounding covariates in treatment $(j=1)$ and control $(j=0)$ groups.
The FDR for these tests were estimated using the method by \cite{ST2003}.

We also consider the case with survival responses.
We generated $n$ independent survival times from the following regression model:
$$
\widetilde{Y}=\exp\left(\sum_{k=1}^p\gamma_kX_k+\sum_{k=1}^p\delta_kX_k^2+T\sum_{k=1}^p\beta_kX_k+\xi_1TZ_1+\xi_2TZ_2+\ep\right),
$$ 
where $\ep\sim N(0,5^2)$ and all the model parameters were the same as those in the previous parts.
The censoring time was generated from the uniform distribution $U(20,60)$, which induced censoring survival time $Y$.
The censoring rate was around $30\%$ in each simulated dataset.
For the simulated dataset, we applied the proposed two ODP methods (ODP-P and ODP-N) with the use of negative log Cox partial likelihood function for the loss function, and the same numbers of knots.
As competitors, we adopted the same form of test statistics as (\ref{standard}), where $\beh_{kj}$ and $s_{kj}$ are obtained from Cox regression with only $k$th biomarker in treatment $(j=1)$ and control $(j=0)$ groups.

Based on 200 simulations, we calculated the average number of significant biomarkers and true positives at FDR$=$5, 10, 15, or 20\% for three types of responses, which are summarized in Table \ref{tab:sim}. 
Overall, the proposed methods detected the larger numbers of biomarkers than the standard methods, which clearly shows the efficiency of the propose methods.
Moreover, the numbers of true positives of the proposed methods are much larger than the standard methods when we allow larger FDR. 
Also the efficiency gain of the proposed method emerged as $\pi$ (null probability) gets smaller.
Comparing the two proposed methods, ODP-P tends slightly more efficient than ODP-N possibly because the approximation error of the normal approximation used in ODP-N may not be negligible and it may reduce the power of ODP-N.
It is also observed that the number of knots has little effect on the results.

%-------------------------------------------------------------%
%    Example
%-------------------------------------------------------------%
\section{Application to a breast cancer clinical study}\label{sec:app}
\begin{figure}[!htb]
\centering
\includegraphics[width=14cm,clip]{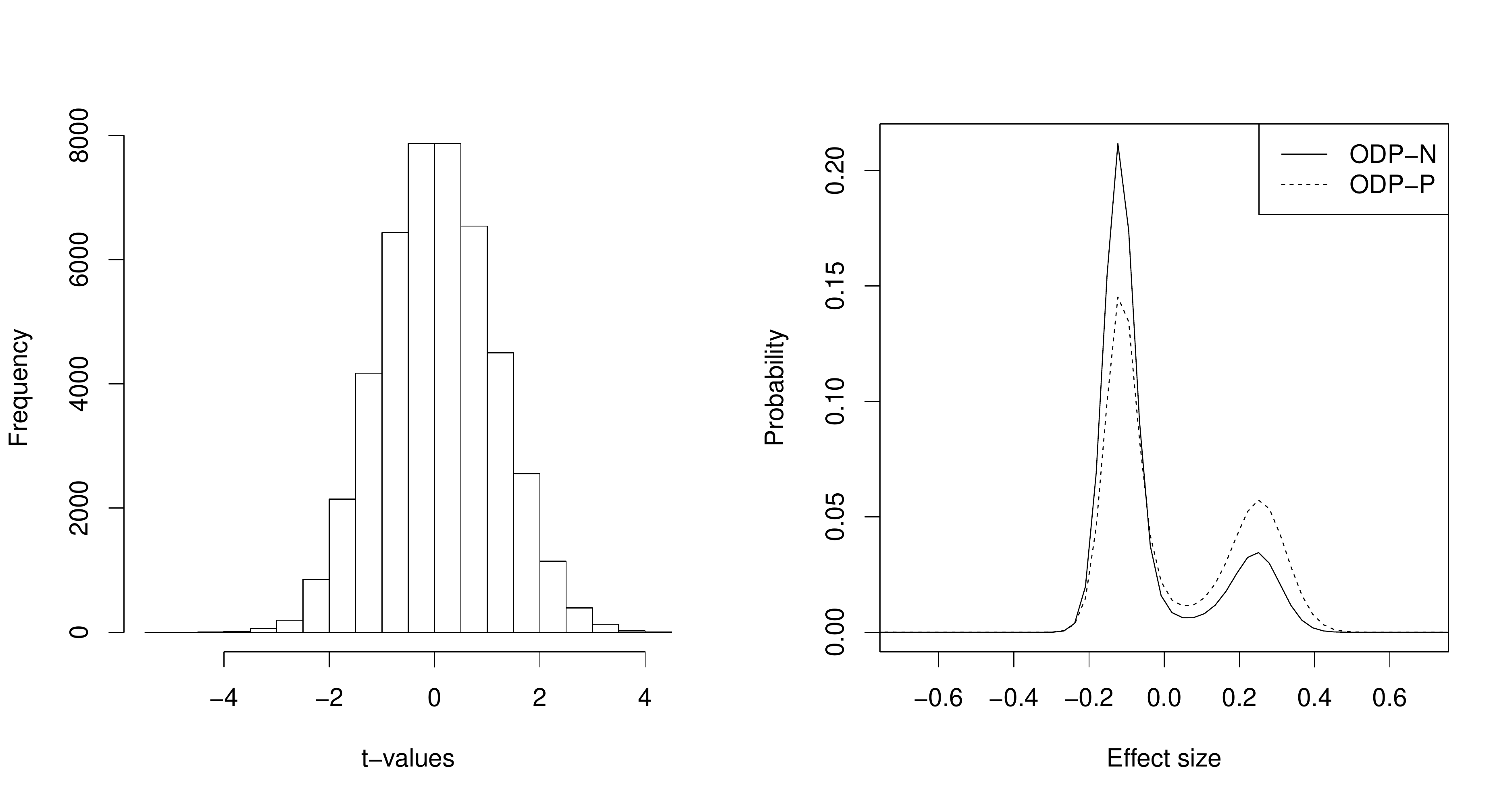}
\caption{Histogram of $t$-values of genes (left) and estimated distribution of underlying effect sizes (right) in breast cancer data.}
\label{fig:Luminal}
\end{figure}

In this section, we illustrate the practical application of the proposed method in a breast cancer clinical study \citep{Loi2007}. 
The dataset consisted of 414 estrogen receptor (ER)-positive breast cancer patients with microarray gene expression profiling data. 
We applied the proposed ODP method to detect the significant gene expression associated with the individual treatment effects. 
The data are available from the NCBI GEO database (GSE 6532). 
Here we adopted the time to relapse-free survival as the outcome. 
After excluding patients with incomplete information, there were 268 and 125 patients receiving tamoxifen and alternative treatments, respectively. 
For each patient, $p=44928$ gene expression measurements were available, dented by $X_1,\ldots,X_p$, and there is no covariates for adjustment.
For estimating the propensity score $\pi(X)$, we employed the following logistic linear regression:
$$
P(T=1|X)=\text{logit}^{-1}\Big(\gamma_0+\sum_{k=1}^p\gamma_kX_k\Big).
$$
We estimated the regression coefficients penalized likelihood with lasso \citep{Tib1996}, where the tuning parameter was selected by 10-fold cross validation.
With use of the estimated propensity score, we applied the two types of proposed ODP methods, ODP-P and ODP-N, with negative log-binomial likelihood and $L=200$ as the number of knots.
The estimated null probabilities were $0.79$ and $0.72$ for ODP-N and ODP-P, respectively, and estimated underlying distributions of effect sizes are shown in the right panel of Figure \ref{fig:Luminal} with histogram of $t$-values of each biomarker. 
From Figure \ref{fig:Luminal}, both methods produced similar estimates of the underlying distribution, and the true underlying distribution of effect sizes would not have a simple form and seem to have two modes.
The numbers of detected microarrays with different FDR values based on the proposed methods and the two standard methods describe in Section \ref{sec:sim} are reported in Table \ref{tab:app}. 
The proposed methods were able to detected much larger numbers of microarrays than the direct use of $q$-values with statistics $T_k$ and $S_k$.
Also, ODP-P found more microarrays than ODP-N possibly because the normal approximation is not accurate in this situation where $n$ is around 400.

%   Discussion
\section{Discussions}\label{sec:dis}
We developed an efficient method for detecting predictive biomarkers associated with individual treatment effects based on the optimal discovery procedure with the synthetic posterior for effect sizes.
The key feature of our approach is to employ the idea of direct estimation of individual treatment effect via the weighted loss function \citep{Chen2017} and combine the semi-parametric prior for interaction effects, which enables flexible and efficient estimation of the underlying signal components.
Employing the empirical Bayes method, the signal components can be accurately estimated from the data.
Then, the estimated model is combined with the optimal discovery procedure to make up a effective screening procedure under which the treatment effects are heterogenous, i.e. there exist predictive biomarkers.
The advantage of the proposed method compared with existing approaches was confirmed through simulation study and the application to breast cancer data.

Although we adopted the independence structure for prior latent distribution for effect sizes of candidate biomarkers, they are typically correlated in practice.
However, the optimality of the optimal discovering procedure is still valid even under potential correlations among effect sizes or candidate biomarkers \cite[e.g.][]{Storey2007,NM2012} and FDR is adequately controlled as long as underlying (marginal) distribution of $\beta_k$ is correctly specified. 
This condition would be naturally satisfied since we employed semiparametric hierarchical mixture modeling. 
Moreover, the underlying distribution of $\beta_{k}$ could be unbiasedly estimated even under potential correlation while the standard error could be larger than that of the model with taking account of potential correlations, as investigated in \cite{MN2011}.
In fact, our simulation study in Section \ref{sec:sim} adopted situations where the biomarkers are correlated, thereby the independence assumption for the effect sizes is violated, but the proposed method performed reasonably well.

Although we employed the weighted loss function \cite{Chen2017} for constructing synthetic posterior distribution (\ref{pos}), other types of loss functions can be readily used in the proposed method by simply changing the formula of synthetic likelihood (\ref{Syn}).
For instance, A-learning method \citep{Murphy2003,Lu2013,Cia2015}, which also can directly estimate individual treatment effect, would be useful alternatives.
Since the detailed investigation and comparison among several types of loss functions would extend the scope of this paper, it is left to a future study.

%  Acknowledgements
\bigskip
\begin{center}
{\bf Acknowledgements}
\end{center}

\noindent
This work was supported by JST CREST Grant Number JPMJCR1412, JST AIP-PRISM Grant Number JPMJCR18Z9 and JSPS KAKENHI Grant Numbers JP17K19808, JP15K15954, JP16H07406, Japan.

%-------------------------------------------------------------%
%    Appendix
%-------------------------------------------------------------%

\vspace{0.5cm}
\appendix
\begin{center}
{\bf Appendix (EM algorithm)}
\end{center}

Let $\phi=(\pi, p_1,\ldots,p_L)$ be the vector of unknown hyperparameters.
We introduce two types of latent variables, $z_k$ and $w_k$ $(k=1,\ldots,p)$, such that $P(z_k=1)=1-P(z_k=0)=\pi$ and $P(w_k=\ell)=p_\ell$.
Then, logarithm of the synthetic likelihood (\ref{like}) can be augmented as  
$$
\sum_{k=1}^p\left[z_k\Big\{\text{PL}_k(0)+\log\pi\Big\}+(1-z_k)\sum_{\ell=1}^LI(w_k=\ell)\Big\{\text{PL}_k(a_\ell)+\log p_{\ell}\Big\}\right].
$$
With current parameter values $p_\ell^{(t)}$ and $\pi^{(t)}$, we have the following objective function in the M-step:
\begin{align*}
Q(\phi|\phi^{(t)})
&=\sum_{k=1}^p\left[\xi_k^{(t)}\log\pi+(1-\xi_k^{(t)})\sum_{\ell=1}^L\eta_{k\ell}^{(t)}\log p_{\ell}\right],
\end{align*}
where 
$$
\xi_k^{(t)}
=\frac{\pi^{(t)} \text{PL}_k(0)}
{\pi^{(t)} \text{PL}_k(0)+(1-\pi^{(t)}) \sum_{\ell=1}^Lp^{(t)}_\ell\text{PL}_k(a_\ell)} \ \ \ {\rm and} \ \ \ 
\eta_{k\ell}^{(t)}
=\frac{p^{(t)}_\ell\text{PL}_k(a_\ell)}{\sum_{j=1}^Lp^{(t)}_j\text{PL}_k(a_j)}.
$$
The updating steps are given by 
\begin{align*}
\pi^{(t+1)}=\frac1p\sum_{k=1}^p\xi_k^{(t)}  \ \ \ {\rm and} \ \ \ 
p_{\ell}^{(t+1)}
=\frac1p\sum_{k=1}^p\eta_{k\ell}^{(t)}, \ \ \ \ \ell=1,\ldots,L.
\end{align*} 
Hence, each iteration in the EM algorithm requires computing $\xi_k^{(t)}$ and $\eta_{k\ell}^{(t)}$ as the E-step and updating $\pi$ and $p_\ell$'s as the M-step, which is continued until numerical convergence.

\vspace{1cm}
%-------------------------------------------------------------%
%    Reference
%-------------------------------------------------------------%

\newpage

%  Table
\begin{table}[htbp]
\caption{Average numbers of true positives and false positives for the proposed methods with difference number of knots ($L$) and alternative $q$-value based methods, based on 200 simulations.
\label{tab:sim}}
\begin{center}
\begin{tabular}{lcccccccccc}
\hline
& \multicolumn{4}{c}{\# true positive} &&\multicolumn{4}{c}{\# false positive} \\
FDR levels & 5\% & 10\% & 15\% & 20\% && 5\% & 10\% & 15\% & 20\%\\
 \hline
Binary ($\pi=0.8$) \\
ODP-P ($L=100$) & 0.83 & 4.91 & 12.59 & 23.44 &  & 0.13 & 1.06 & 3.76 & 9.56 \\
ODP-N ($L=50$) & 0.33 & 2.24 & 7.57 & 17.00 &  & 0.04 & 0.42 & 1.89 & 6.13 \\
ODP-N ($L=100$) & 0.33 & 2.25 & 7.58 & 17.07 &  & 0.04 & 0.43 & 1.90 & 6.15 \\
ODP-N ($L=150$) & 0.33 & 2.25 & 7.59 & 17.09 &  & 0.04 & 0.43 & 1.90 & 6.15 \\
ODP-N ($L=200$) & 0.33 & 2.25 & 7.59 & 17.10 &  & 0.04 & 0.43 & 1.90 & 6.15 \\
$T_k$ & 0.40 & 1.20 & 2.33 & 3.91 &  & 0.07 & 0.29 & 0.71 & 1.28 \\
$S_k$ & 0.35 & 1.11 & 2.08 & 3.44 &  & 0.07 & 0.27 & 0.56 & 1.05 \\
\hline
Binary ($\pi=0.5$) \\
ODP-P ($L=100$) & 0.36 & 4.80 & 18.47 & 45.64 &  & 0.07 & 0.82 & 4.25 & 12.67 \\
ODP-N ($L=50$) & 0.22 & 2.14 & 12.45 & 39.36 &  & 0.03 & 0.34 & 2.77 & 10.58 \\
ODP-N ($L=100$) & 0.22 & 2.14 & 12.51 & 39.53 &  & 0.03 & 0.34 & 2.79 & 10.63 \\
ODP-N ($L=150$) & 0.22 & 2.14 & 12.53 & 39.57 &  & 0.03 & 0.34 & 2.79 & 10.65 \\
ODP-N ($L=200$) & 0.22 & 2.14 & 12.54 & 39.59 &  & 0.03 & 0.34 & 2.79 & 10.66 \\
$T_k$ & 0.30 & 0.71 & 1.49 & 2.85 &  & 0.04 & 0.13 & 0.32 & 0.63 \\
$S_k$ & 0.29 & 0.71 & 1.31 & 2.33 &  & 0.03 & 0.13 & 0.26 & 0.51 \\
\hline
Survival ($\pi=0.8$) \\
ODP-P ($L=100$) & 2.86 & 11.45 & 24.35 & 40.81 &  & 0.34 & 2.20 & 6.68 & 15.90 \\
ODP-N ($L=50$) & 2.22 & 9.53 & 21.87 & 39.03 &  & 0.24 & 1.91 & 6.43 & 15.54 \\
ODP-N ($L=100$) & 2.23 & 9.54 & 21.93 & 39.12 &  & 0.24 & 1.92 & 6.44 & 15.60 \\
ODP-N ($L=150$) & 2.23 & 9.55 & 21.97 & 39.15 &  & 0.24 & 1.92 & 6.44 & 15.61 \\
ODP-N ($L=200$) & 2.23 & 9.55 & 21.98 & 39.17 &  & 0.24 & 1.92 & 6.44 & 15.62 \\
$T_k$ & 1.35 & 3.35 & 6.39 & 10.51 &  & 0.17 & 0.67 & 1.64 & 3.56 \\
$S_k$ & 1.16 & 2.80 & 5.66 & 9.74 &  & 0.15 & 0.50 & 1.29 & 2.93 \\
\hline
Survival ($\pi=0.5$) \\
ODP-P ($L=100$) & 1.34 & 11.84 & 36.93 & 79.33 &  & 0.13 & 1.92 & 7.91 & 22.08 \\
ODP-N ($L=50$) & 0.95 & 8.43 & 30.75 & 76.12 &  & 0.11 & 1.43 & 7.21 & 21.92 \\
ODP-N ($L=100$) & 0.96 & 8.47 & 30.85 & 76.31 &  & 0.11 & 1.44 & 7.23 & 22.02 \\
ODP-N ($L=150$) & 0.96 & 8.49 & 30.88 & 76.38 &  & 0.11 & 1.45 & 7.24 & 22.03 \\
ODP-N ($L=200$) & 0.96 & 8.50 & 30.90 & 76.40 &  & 0.11 & 1.45 & 7.24 & 22.04 \\
$T_k$ & 0.64 & 1.67 & 3.12 & 6.04 &  & 0.08 & 0.24 & 0.63 & 1.30 \\
$S_k$ & 0.57 & 1.46 & 2.92 & 5.07 &  & 0.06 & 0.24 & 0.54 & 1.10 \\
\hline
\end{tabular}
\end{center}
\end{table}

%%  Table
%\begin{table}[htbp]
%\caption{Average numbers of true positives and false positives for the proposed methods with difference number of knots ($L$) and alternative $q$-value based methods, based on 200 simulations.
%\label{tab:sim}}
%\begin{center}
%\begin{tabular}{lcccccccccc}
%\hline
%& \multicolumn{4}{c}{\# true positive} &&\multicolumn{4}{c}{\# false positive} \\
%FDR levels & 5\% & 10\% & 15\% & 20\% && 5\% & 10\% & 15\% & 20\%\\
%\hline
%ODP-P  & 2.86 & 11.45 & 24.35 & 40.81 &  & 0.34 & 2.20 & 6.68 & 15.90 \\
%ODP-N  & 2.23 & 9.54 & 21.93 & 39.12 &  & 0.24 & 1.92 & 6.44 & 15.60 \\
%$T_k$ & 1.35 & 3.35 & 6.39 & 10.51 &  & 0.17 & 0.67 & 1.64 & 3.56 \\
%$S_k$ & 1.16 & 2.80 & 5.66 & 9.74 &  & 0.15 & 0.50 & 1.29 & 2.93 \\
%\hline
%ODP-P  & 1.34 & 11.84 & 36.93 & 79.33 &  & 0.13 & 1.92 & 7.91 & 22.08 \\
%ODP-N  & 0.96 & 8.47 & 30.85 & 76.31 &  & 0.11 & 1.44 & 7.23 & 22.02 \\
%$T_k$ & 0.64 & 1.67 & 3.12 & 6.04 &  & 0.08 & 0.24 & 0.63 & 1.30 \\
%$S_k$ & 0.57 & 1.46 & 2.92 & 5.07 &  & 0.06 & 0.24 & 0.54 & 1.10 \\
%\hline
%\end{tabular}
%\end{center}
%\end{table}

%  Table
\begin{table}[htbp]
\caption{The numbers of detected genes in breast cancer data.
\label{tab:app}}
\begin{center}
\begin{tabular}{ccccccc}
\hline
FDR levels & 5\% & 10\% & 15\% & 20\%\\
 \hline
ODP-P & 11 & 45 & 104 & 197\\
ODP-N & 10 & 40 & 81 & 140\\
$T_k$ & 1 & 2 & 2 & 2\\
$S_k$ & 0 & 0 & 0 & 0\\
\hline
\end{tabular}
\end{center}
\end{table}

%  Figure

\end{document}